\documentclass{article}
\usepackage{spconf,graphicx}
\usepackage{amssymb,amsmath}
\usepackage{diagbox}
\usepackage{makecell}
\usepackage{booktabs}
\usepackage{enumitem}
\setlist{nosep, leftmargin=14pt}

\usepackage{mwe} 

\begin{document}
\title{
 MD-IQA : Learning 
Multi-scale Distributed Image Quality Assessment with Semi Supervised Learning for
Low Dose CT}
%

%
\name{Tao Song$^{1,2,\dagger}$\thanks{$^{\dagger}$Corresponding author: Tao Song (tsong22@m.fudan.edu.cn)}, Ruizhi Hou$^{1,3}$, Lisong Dai$^{4}$, Lei Xiang$^{1}$ }

\address{
$^{1}$ Subtle Medical, Shanghai, China \\
     $^{2}$ School of Information and Technology, Fudan University, Shanghai, China \\
    $^{3}$ School of Mathematical Science, East China Normal University, Shanghai, China\\
    $^{4}$ Institute of Diagnostic and Interventional Radiology, Shanghai Sixth People's Hospital, Shanghai, China\\
    }



%
\maketitle
\begin{abstract}
Image quality assessment (IQA) plays a critical role in optimizing radiation dose and developing novel medical imaging techniques in computed tomography (CT). Traditional IQA methods relying on hand-crafted features have limitations in summarizing the subjective perceptual experience of image quality. Recent deep learning-based approaches have demonstrated strong modeling capabilities and potential for medical IQA, but challenges remain regarding model generalization and perceptual accuracy. In this work, we propose a multi-scale distributions regression approach to predict quality scores by constraining the output distribution, thereby improving model generalization. Furthermore, we design a dual-branch alignment network to enhance feature extraction capabilities. Additionally, semi-supervised learning is introduced by utilizing pseudo-labels for unlabeled data to guide model training. Extensive qualitative experiments demonstrate the effectiveness of our proposed method for advancing the state-of-the-art in deep learning-based medical IQA. Code is available at: https://github.com/zunzhumu/MD-IQA.
\end{abstract}
\begin{keywords}
image quality assessment, multi-scale distributions regression, semi-supervised learning
\end{keywords}
\section{Introduction}
\label{sec:intro}
Computed tomography (CT) plays an indispensable role in modern medical imaging and diagnostics. However, the radiation exposure in CT scans may raise major health concerns, as high doses of radiation can lead to harmful damage in patients. Low-dose CT has thus emerged as an important technique to mitigate radiation risks. Nonetheless, the reduced radiation in low-dose CT can compromise image quality and impede clinical diagnosis due to increased noise. Therefore, developing accurate IQA methods is crucial for optimizing CT radiation dosages according to diagnostic needs. While conventional metrics like peak signal-to-noise ratio (PSNR) and structural similarity index (SSIM) can evaluate image quality, they rely solely on pixel differences without considering perceptual quality and diagnostic impact. To truly enable dose optimization in CT, it is essential to design IQA techniques that can predict perceptual image quality aligned with radiologists' assessments \cite{NRLDCTIQA1,IQAISBI,CTIQA_SPIE}.

Although recent years have witnessed remarkable progress in IQA methods for natural images \cite{NRIQA,FRIQA,TReS, DBCNN}, IQA research tailored for computed tomography (CT) remains relatively limited. State-of-the-art natural IQA techniques utilizing convolutional and vision transformer networks have demonstrated promising performance, such as the multi-dimension attention network (MANIQA) \cite{NRIQA} and attention-based hybrid IQA model (AHIQ) \cite{FRIQA}. Developing effective CT IQA solutions poses unique challenges. First, the complex CT acquisition pipeline creates artifacts hard to simulate. Second, acquiring pristine CT scans as ground truth is difficult. Third and critically, CT image quality depends on diagnostic value rather than pixel metrics. The lack of unified perceptual standards makes large-scale CT IQA dataset creation costly and unreliable.


\renewcommand{\dblfloatpagefraction}{.9}
\begin{figure*}
\begin{center}
\includegraphics[scale=0.13]{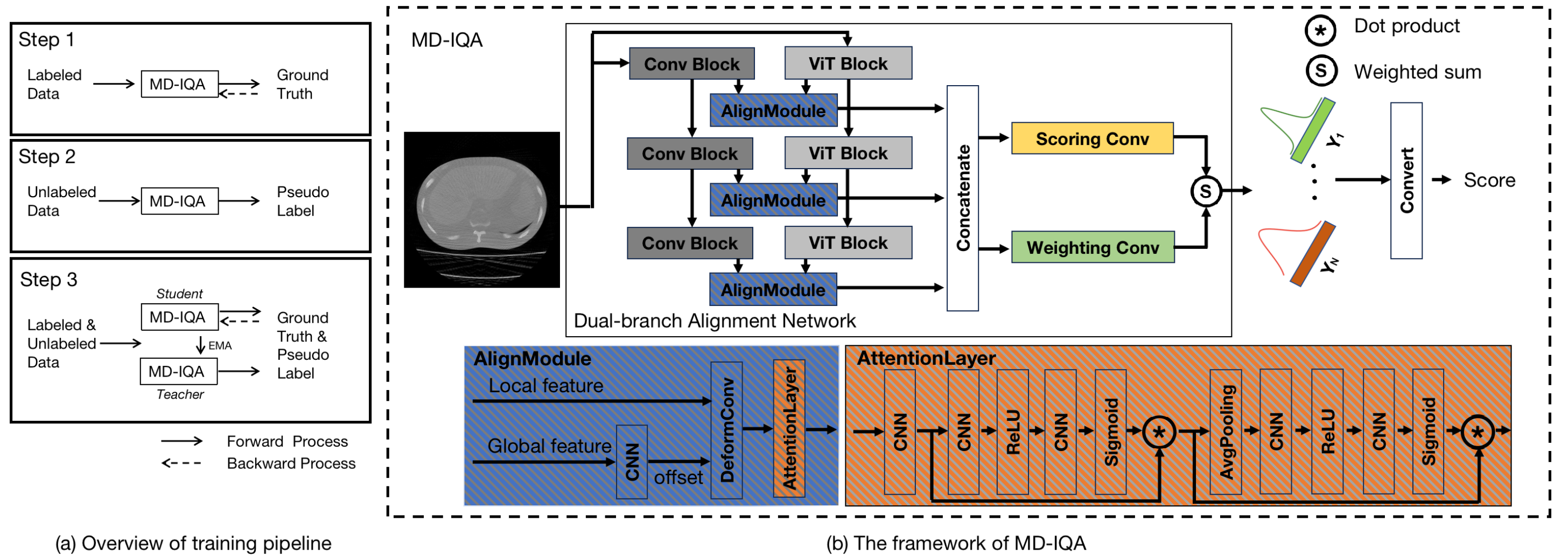}
\end{center}
\caption{The proposed training pipeline and framework for low dose CT image quality assessment. (a) Overview of training pipeline with semi supervised learning. (b) The dual-branch alignment network of our MD-IQA combines transformer and convolution networks through alignment modules to generate multi-scale Gaussian distributions for predicting image quality scores.}
\label{overview}
\end{figure*}

Recent studies on  CT IQA have utilized gradient SSIM \cite{IQAISBI} and object detection performance \cite{NRLDCTIQA1} as surrogate quality measures. However, these metrics lack direct clinical relevance. To enable more perceptually aligned CT IQA, researchers have compiled a low-dose CT (LDCT) dataset \cite{Dataset} with quality scores obtained by averaging five radiologists' subjective ratings. Building upon this preliminary dataset, we aim to develop a regression model for predicting radiologist-judged quality scores. Two key challenges arise in utilizing this data. First, substantial label uncertainty exists due to subjective inter-rater variability and intra-rater inconsistency over time. Therefore, directly regressing the scores using deep learning models will face significant uncertainty in prediction results. Second, the limited number of scored images risks overfitting.

To tackle these issues, we propose a multi-scale distributions regression approach for LDCT IQA called MD-IQA, inspired by heat map regression techniques \cite{HMHumanPose, qu2022heatmap}. Our key contributions are threefold:
First, we introduce multi-scale distributions regression to reduce prediction uncertainty and improve robustness by modeling score distributions instead of regressing point estimates. Second, to further enhance representation capability, we develop a dual-branch alignment network with Vision Transformer (ViT) and CNN modules to extract global and local features respectively. We utilize deformable CNNs to align multi-scale CNN and ViT features. Third, due to the small number of training samples, we adopt semi-supervised learning to train the network on an expanded dataset and improve generalization.



\section{Methods}
\subsection{Problem Formulation}

Semi-supervised learning of IQA aims to learn a more robust network from both the labeled and unlabeled data. The problem  of IQA is defined as follows. 
Let $D_{L} = \left \{ \left ( x_{i}^{L},s_{i} \right )|x_{i}^{L}\in X^{L}, s_{i}\in S  \right \} _{i=1}^{N}$ denote the labeled dataset, where $x_{i}^{L}$ and $s_{i}$ are respectively the labeled LDCT image and corresponding image quality score. Similarly, let $D_{U} = \left \{ (x_{i}^{U}, p_{i}) |x_{i}^{U}\in X^{U}, p_{i}\in P \right \} _{i=1}^{N}$ denote the unlabeled dataset, where $x_{i}^{U}$ and $p_{i}$ are respectively the unlabeled LDCT image and corresponding pseudo image quality score. The pseudo image quality score is generated by a pre-trained model.

\subsection{Multi-scale Gaussian Distribution Regression}
The network is expected to predict several Gaussian distribution with different scales. If the image quality score is $s_{i}\in [0,M]$, the $j$-th element of the corresponding probability density function (PDF) vector $\boldsymbol{Y}[j](\sigma)\in \mathbb{R}^{N}$ of Gaussian distribution satisfies:
\begin{equation}\label{eq:hm_define}
\boldsymbol{Y}[j](\sigma) = \text{exp}(-\frac{(j - u)^2}{2\sigma^2})\\ 
\end{equation}
here $u = (s_{i}/M)*N$, $\sigma = N * \tau$, where $M$ is the maximum score and $\tau$ is the scale of Gaussian distribution. The sampling point number of the Gaussian distribution vector is denoted as $N$, and a larger $N$ indicates the vector can represent a more accurate score distribution. During inference, the predicted score can be calculated as $\tilde{p}(\sigma) = \frac{argmax(Y(\sigma))}{N} * M$.
The predicted multi-scale Gaussian distribution scores can be formulated as $
p=\frac{1}{K} \sum_{i}^{K} \tilde{p}(\sigma_{i})$, where $K$ denotes the number of distinct scales for the Gaussian distributions.

According to the PDF curve of the Gaussian distribution, the predicted values near the center position are relatively large. Therefore, the network first roughly chooses a region covering the possible score values, and then predicts the final score. In this way, the Gaussian distribution allows for uncertainty in the label. However, a single-scale Gaussian distribution cannot handle varying degrees of uncertainty. Using multi-scale Gaussian distributions can better measure the label uncertainty. Therefore, using multi-scale Gaussian distribution regression can achieve more robust and accurate performance in the IQA task for LDCT.


\subsection{Dual-branch Alignment Network}
The proposed network consists of three key components: a feature extraction module, dual branch align module, and a score prediction module, as depicted in the overall architecture in Fig.1 (b).

The feature extraction module consists of two submodules: ViT \cite{ViT} and ConvNeXt module \cite{ConvNeXt}. The ViT module splits the image into patches and encodes them into feature representations, then applies self-attention layers to extract global information from the feature patches. The ConvNeXt module serves as the convolutional submodule to extract local features. In the align module, we perform alignment using deformable convolution, resizing the multiple CNN feature maps to match the ViT feature size. The ViT features are input to a CNN layer to predict sampling offsets for deformable convolution over the CNN features. Attention layers are then applied to the aligned CNN features.

The final module predicts the overall quality score by weighted summation of the fused features. Since each pixel in the extracted feature maps represents an image patch embedding, we first predict quality scores for individual patches, then take a weighted average to obtain the score for the whole image.
Specifically, let $\boldsymbol{f}\in\mathbb{R}^{C\times H\times W}$ denote the fused feature maps. We compute patch-wise scores $\boldsymbol{f'}$ by applying a $Scoring$ $Conv$ over $\boldsymbol{f}$. The weight $\boldsymbol{w}$ is obtained by a sigmoid activation on $\boldsymbol{f'}$, which applying a  $Weighting$ $Conv$. For a image, the overall quality score distributions vector $\boldsymbol{Y}$ is then calculated as:

\begin{equation}
    \boldsymbol{Y} = \frac{\sum_{i=1}^{W}\sum_{j=1}^{H} \boldsymbol{w}[:,i,j]\odot \boldsymbol{f'}[:,i,j]}{\sum_{i=1}^{W}\sum_{j=1}^{H} \boldsymbol{w}[:,i,j]},
\end{equation}
where $\odot$ is the element-wise product.

\subsection{Training Pipeline}
The training pipeline (Fig 1. (a)) consists of 3 steps as follows:
\begin{itemize}
    \item[1] \textbf{Supervised Learning.} We train several MD-IQA models with different training settings on the training dataset.
    \item[2] \textbf{Pseudo Labeling.} We generate the pseudo labels for the unlabeled extended dataset using the ensemble models trained in step 1. 
    \item[3] \textbf{Jointly training.} We use the datasets with true and pseudo labels to train a new MD-IQA model.
\end{itemize}

In step 3, the datasets containing images with true labels and pseudo labels are used to train the model jointly. During training, the image-label pairs consist of $(\boldsymbol{X}^{L}, \boldsymbol{\hat{Y}}) \in \mathcal{D}_{L}$ from the labeled dataset and $(\boldsymbol{X}^{U}, \boldsymbol{P}) \in \mathcal{D}_{U}$ from the unlabeled dataset. The overall loss function $\mathcal{L}_{total}$ in step 3 is defined as:
\begin{equation}
\label{eq:loss_function_sum}
\mathcal{L}_{total}
=  \lambda_1{L_{sup}} 
+ \lambda_2{L_{pseudo}}
+ \lambda_3{L_{consistency}} 
\end{equation}
\begin{equation}
{L_{sup}} = L_2(\phi, \boldsymbol{\hat{Y}}) + \beta{L}_{KL}(\phi, \boldsymbol{\hat{Y}}), \phi = F(\boldsymbol{X}^{L};\boldsymbol{\Theta}^{(t)})
\end{equation}
\begin{equation}
L_{pseudo}= L_2(\xi, \boldsymbol{P})+ \beta{L}_{KL}(\xi, \boldsymbol{P}),\xi=F(\boldsymbol{X}^{U};\boldsymbol{\Theta}^{(t)})
\end{equation}
\begin{equation}
 L_{consistency} = L_2(F(\boldsymbol{X}^{U};\boldsymbol{\Theta}^{(t)}), F((\boldsymbol{X}^{U};\boldsymbol{\Theta}_{EMA}^{(t)}))
\end{equation}
where $F$ is the network, $L_{KL}$ is the Kullback-Leibler (KL) divergence between the predicted distribution and target distribution, $\boldsymbol{\Theta}^{(t)}$ represents the network parameters at the $t$-th iteration, and $\boldsymbol{\Theta}_{EMA}^{(t)}$ is the exponential moving average (EMA) of the parameters from previous iterations. 
The EMA weights $\boldsymbol{\Theta}_{EMA}^{(t)}$ are updated as $\boldsymbol{\Theta}_{EMA}^{(t+1)}=\alpha\boldsymbol{\Theta}_{EMA}^{(t)}+(1-\alpha)\boldsymbol{\Theta}^{(t)}$, where $\alpha$ is the decay parameter. During supervised learning in step 1, only the supervised loss $L_{sup}$ is used to optimize the MD-IQA model weights.



\section{Experiments and Results}
\subsection{Data and Implementation}
The LD-CT IQA dataset\footnote{https://ldctiqac2023.grand-challenge.org} \cite{Dataset} contains 1,000 CT image slices exhibiting different levels of aliasing artifacts and noise caused by low-dose radiation. The image quality of each CT slice was scored by five experienced radiologists using abdominal soft-tissue windows, with scores ranging from 0 to 4. A score of 0 indicates the worst quality, while 4 indicates the best quality. The final human perceptual score for each image was calculated by averaging the scores from the five radiologists. The images have a resolution of  $512\times 512$ pixels. We divided the dataset into training, validation, and testing subsets containing 700, 100, and 200 images respectively. Additionally, we collected an extended dataset of 700 real-world LD-CT images with no ground truth quality scores.

Data augmentation is performed by randomly rotating and flipping the images. The feature extraction module is initialized with pre-trained weights from ViT and ConvNeXt models. During training, the batch size is set to 16 and the number of epochs is 30. The Adam optimizer is used to update network weights, with a learning rate of 0.0001. The EMA decay parameter $\alpha$ is set to 0.997, supervised loss weight $\lambda_1$ to 1, unsupervised loss weight $\lambda_2$ to 0.1, and KL constraint weight $\beta$ to 0.1. A warmup strategy gradually increases $\lambda_3$ as training progresses, such that $\lambda_3$ surpasses $\lambda_2$ after 400 iterations and approaches 1. All experiments are implemented in PyTorch and run on NVIDIA GeForce RTX 3090 GPUs.

\begin{table}[t]
\caption{Quantitative results of compared methods.}
\label{Table:results}
\centering
\resizebox{\linewidth}{!}{
\begin{tabular}{ccccc}
\toprule
\diagbox{Methods}{Metrics} & $|PLCC|$ $\uparrow$ & $|SROCC|$ $\uparrow$ & $|KROCC|$ $\uparrow$ & $Overall$ $\uparrow$\\ \midrule
DBCNN \cite{DBCNN}             & 0.9725 & 0.9723 & 0.8742 & 2.8189  \\
TReS \cite{TReS}             & 0.9755 & 0.9745 & 0.8786 & 2.8286  \\
AHIQ \cite{FRIQA}             & 0.9762 & 0.9746 & 0.8810 & 2.8317  \\
MANIQA \cite{NRIQA}           & 0.9768 & 0.9786 & 0.8891 & 2.8445  \\
Ours               & \textbf{0.9771} & \textbf{0.9793} & \textbf{0.9106} & \textbf{2.8670} \\
\bottomrule
\end{tabular}
}
\end{table}

\begin{table}[t]
\caption{Ablation study. MD: Multi-scale distributions regression. SS: Semi-supervision.}
\label{Table:ablation}
\centering
\resizebox{\linewidth}{!}{
\begin{tabular}{cc|cccc}
\toprule
\multicolumn{2}{c|}{Methods} & \multicolumn{4}{c}{Metrics} \\ \midrule
MD & SS & $|PLCC|$ $\uparrow$ & $|SROCC|$ $\uparrow$ & $|KROCC|$ $\uparrow$ & $Overall$ $\uparrow$ \\ \midrule
 & & 0.9751          & 0.9758          & 0.8846          & 2.8355 \\
 & $\checkmark$ & 0.9774          & 0.9765         & 0.8872          & 2.8411 \\ 
$\checkmark$ & & 0.9757          & 0.9755          & 0.9020          & 2.8532 \\
$\checkmark$ & $\checkmark$ & \textbf{0.9771} & \textbf{0.9793} & \textbf{0.9106} & \textbf{2.8670} \\
\bottomrule
\end{tabular}
}\end{table}

\begin{figure}
\begin{center}
\includegraphics[width=\linewidth, scale=0.6]{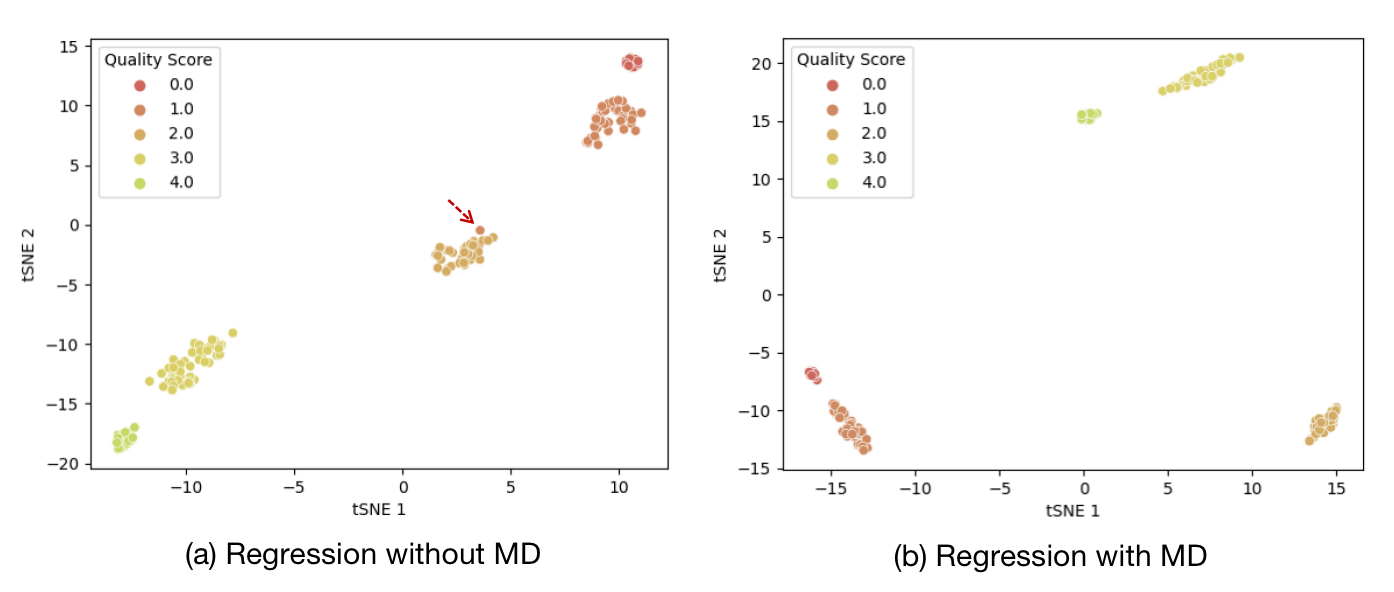}
\end{center}
\caption{Visualization of hidden layer feature representations using t-SNE. Features were extracted from the multi-scale Gaussian distribution regression model (left) and direct regression model (right). The multi-scale distribution features show more distinct clustering compared to the direct regression features.}
\label{overview}
\end{figure}


\begin{table}[t]
\caption{Quantitative comparison of the top five methods for LDCTIQAC2023 testing set.}
\label{Table:results}
\centering
\resizebox{\linewidth}{!}{
\begin{tabular}{ccccc}
\toprule
\diagbox{Methods}{Metrics} & $|PLCC|$ $\uparrow$ & $|SROCC|$ $\uparrow$ & $|KROCC|$ $\uparrow$ & $Overall$ $\uparrow$\\ \midrule
$1^{st}$            & 0.9497 & 0.9502 & \textbf{0.8456} & \textbf{2.7455}  \\
Ours              & \textbf{0.9513} & \textbf{0.9514} & 0.8404 & 2.7431  \\
$3^{rd}$             & 0.9471 & 0.9470 & 0.8340 & 2.7282  \\
$4^{th}$           & 0.9394 & 0.9414 & 0.8248 & 2.7057  \\
$5^{th}$              & 0.9485 & 0.9465 & 0.8081 & 2.7031 \\
\bottomrule
\end{tabular}
}
\end{table}

\subsection{Metrics}
The evaluation metrics from LDCTIQAC2023\footnote{https://ldctiqac2023.grand-challenge.org} used to measure the accuracy of the proposed algorithm are absolute Pearson Linear Correlation Coefficient ($|PLCC|$), absolute Spearman rank-order correlation coefficient ($|SROCC|$), and absolute Kendall rank-order correlation coefficient ($|KROCC|$). Given two vectors $\boldsymbol{x},\boldsymbol{y}\in\mathcal{R}^{N}$, $|PLCC|$ is defined as:
\begin{equation}
    |PLCC(\boldsymbol{x},\boldsymbol{y})| = |\frac{\sum_{i=1}^{n}({x}_i-\bar{x})(y_i-\bar{y})}{\sqrt{{\sum_{i=1}^{n}(x_i-\bar{x})^2}{\sum_{i=1}^{n}(y_i-\bar{y})^2}}}|
\end{equation}
where $\bar{x}=\frac{1}{n}\sum_{i=1}^{n}x_{i}$ and $\bar{y}=\frac{1}{n}\sum_{i=1}^{n}y_{i}$.
Similarly, $|SROCC|$ is defined as:
\begin{equation}
    |SROCC(\boldsymbol{x},\boldsymbol{y})| = |1 - \frac{6\sum_{i=1}^{n}d_{i}^2}{n(n^2-1)}|
\end{equation}
where $d_{i}$ is the difference between the $i^{th}$ image's ranks in the subjective and objective evaluations. Finally, $|KROCC|$ is defined as:
\begin{equation} 
|KROCC(\boldsymbol{x},\boldsymbol{y})| = |(C-D)/ \binom{n}{2}|
\end{equation}
where C stands for the number of concordant pairs between $\boldsymbol{x}$ and $\boldsymbol{y}$, while D denotes the number of discordant pairs. So the $Overall$ is defines as $Overall=|PLCC|+|SROCC|+|KROCC|$.




\subsection{Results}
We compared our proposed method to several existing IQA methods, including MANIQA \cite{NRIQA} and AHIQ \cite{FRIQA}, which were the 1st place solutions in the NTIRE IQA challenge full-reference and no-reference tracks, respectively. These methods focus on network architecture design. We also compared to DBCNN\cite{DBCNN} and TReS\cite{TReS}, which concentrate on training strategies and loss functions. The quantitative results comparing these methods are presented in Table 1. Our method achieved average $|PLCC|$, $|SROCC|$, $|KROCC|$ and $Overall$ scores of 0.9771, 0.9793, 0.9106 and 2.8670, respectively. These results are significantly higher than the other four benchmark methods.

We conducted ablation studies to investigate the effectiveness of semi-supervision and multi-scale Gaussian distribution (MD) regression. As shown in Table 2, adding semi-supervision improves performance over the baseline model without MD regression. Comparing the baseline to the MD regression model shows that MD regression provides further gains. Although our task involves regressing a quality score, the ground truth values are evenly distributed between 0 and 4 at 0.2 intervals. To demonstrate the benefits of MD regression, we used t-SNE to visualize the hidden features learned with and without MD regression (Fig. 2). The features from MD regression are learned more accurately, with more compact clustering. Finally, we applied semi-supervision to the MD regression model. This further improved performance compared to MD regression alone

We also compared our method to the top-performing methods on the LDCTIQAC2023 testing set, as shown in Table 3. Our method achieved second place on the testing phase leaderboard \footnote{https://ldctiqac2023.grand-challenge.org/evaluation/training/leaderboard/}, outperforming the first place method on both $|PLCC|$ and $|SROCC|$ metrics. These quantitative results demonstrate the generalization ability of our approach on unseen data.

\section{Discussion and Conclusion}

In this work, we proposed a multi-scale distribution regression model with semi-supervised learning (MD-IQA) for accurately assessing image quality of low-dose CT images. The multi-scale distributions help the network learn more robust and accurate representations. Pseudo-labeling further improves the performance of MD-IQA under limited labeled data. Comprehensive experiments on the public LDCTIQAC2023 challenge dataset validate the effectiveness of our proposed approach. MD-IQA outperforms several existing natural image quality assessment methods when evaluated on low-dose CT images. In the future, it would be interesting to apply and evaluate our method on low dose CT reconstruction task.
\section{COMPLIANCE WITH ETHICAL STANDARDS}
This research study was conducted retrospectively using human subject data made available in open access\cite{Dataset}. Ethical approval was not required as confirmed by the license attached with the open access data.
\bibliographystyle{IEEEbib}
\bibliography{refs}

\end{document}